\documentclass[12pt]{iopart}
\usepackage{float} 
\usepackage{graphicx}
\begin{document}

\title[Remarks on projected densities]{Remarks on the use
of projected densities in the density dependent part of  
Skyrme or Gogny functionals}

\author{L M Robledo}

\address{Departamento de F\'\i sica Te\'orica (M\'odulo 15),
Universidad Aut\'onoma de Madrid, 
28049 Madrid, Spain}
\ead{luis.robledo@uam.es}
\begin{abstract}
I discuss the inadequacy of the  ``projected density" prescription to be used
in density dependent forces/functionals when calculations beyond mean field are pursued. 
The case of calculations aimed at the symmetry restoration of mean fields obtained with 
effective realistic forces of the Skyrme or Gogny type is considered in detail. It is shown 
that at least for the restoration of spatial symmetries like rotations, translations or 
parity the above prescription yields catastrophic results for the
energy that drive the intrinsic wave function to configurations with infinite deformation,
preventing thereby its use both in projection after and before variation.

\end{abstract}

\submitto{\JPG}
\maketitle

\section{Introduction}

Given the enormous success of phenomenological density dependent forces/functionals like the
different variants of Skyrme, Gogny or relativistic mean fields  in describing
nuclear properties with mean field wave function \cite{Bender.03}, it was expected that going 
beyond mean field both restoring broken symmetries or considering configuration mixing 
would pave the way for a more satisfactory understanding of nuclear structure both 
in the qualitative and quantitative sides. Early attempts with the Skyrme functional 
\cite{Bonche.90} applying the Generator Coordinate Method (GCM) produced very encouraging 
results. Soon after, fully microscopic parity projection
was implemented with the Gogny force \cite{Egido.91} for octupole deformed states 
in a consistent framework and the comparison with other results as well as with
experiment was encouraging. After that, many GCM and symmetry restoration calculations
both with Skyrme and Gogny were reported. Here, I will only mention a few like 
the restoration of particle number  symmetry that was tackled in \cite{Anguiano.01} 
with the Gogny force. In this reference it was realized that a 
consistent inclusion of the direct, exchange and pairing terms of the interaction was
needed for a divergent free evaluation of the energy. This was a real problem for 
functionals of the Skyrme type were the exchange and pairing fields of different terms of 
the functional were usually neglected  (probably the most disappointing situation was the
one of the Coulomb potential, as the
consideration of its exchange and pairing fields increased by orders of magnitude
the computational cost of the Skyrme type calculations). Recently, a regularization 
procedure to solve this problem and also to address 
the inherent self-energy and self-pairing problem of any energy density functional was proposed 
\cite{Lacroix.09} and applied to the case of particle number projection (PNP) in \cite{Bender.09}.
Although the regularization procedure of \cite{Lacroix.09} is basis-dependent and the
impact of this arbitrariness has still to be checked, the results obtained so far \cite{Bender.09}
are very promising. 

Another problem that plagues beyond mean field calculations has to do with the
prescription to be used for the density dependent part of the ``Hamiltonian"/functional
(quotes will be  used through the paper to stress the dependency of the ``Hamiltonian" operator 
with the intrinsic wave function through the density dependent part of the interaction). The
first prescription used \cite{Bonche.90} was inspired by the three body force origin of some Skyrme
functionals and is referred to as the ``mixed density" prescription (see below). Although 
attempts to use other prescriptions are not uncommon 
\cite{Duguet.03,Valor.97,Valor.00,Schmid.04} it was assumed that the consistency requirements 
summarized in \cite{Robledo.07} favored the use of the ``mixed density" prescription. However,
as pointed out in \cite{Dobaczewski.07,Duguet.09} the ``mixed density" prescription has a serious
and in general insurmountable drawback when the density dependent part of the interaction 
depended upon non-integer powers of the density. As the mixed density is in general a
complex quantity, non integer powers of it are not uniquely defined in the complex plane unless
a ``branch cut" is introduced and the Riemann sheet where the root lives is specified. 
In many cases, the mixed density is a real
quantity and such a problem can be circumvented, but in general (including the most complex 
symmetry restoration of triaxial deformed states) it is complex and the lack of a physically 
guided rule to choose the roots yields to projected energies that depend on external artifacts 
like the integration contours entering the definition of the projector operators \cite{Dobaczewski.07,Duguet.09}. 
The only two known ways out to this problem in the general case are either to consider forces/functionals 
depending upon integer powers of the density (see below for a recently proposed functional) or 
to consider other prescriptions producing real densities. In the latter case, the most 
promising candidate is the ``projected density" prescription \cite{Valor.97,Valor.00,Schmid.04}
that, in addition, has some other 
appealing properties to be discussed below. In this paper, I am going to show by means of an
example concerning parity projection that the ``projected density" prescription yields to 
unphysical energies that favor infinitely deformed intrinsic states with diverging energies.
This catastrophic behavior can also be extended to other spatial symmetry restorations like
angular momentum projection or translational invariance. 

I would like to emphasize that the present discussion concerns the
inadequacy of using on the same footing symmetrized densities and
intrinsic ones to obtain the energy associated to the strongly 
repulsive density dependent interaction,
which is characteristic of the nuclear Energy Density Functional (EDF).
The present discussion has little to do with the solution 
\cite{Gorling.93} in terms of symmetrized density matrices
of the symmetry dilemma of the Kohn-Sham Density Functional Theory (DFT).
The symmetry dilemma refers to the use of KS-DFT density functionals
which break in some circumstances the symmetries of the interaction
and lead to Kohn-Sham wave functions that, in contrast to the
full interacting wave function, cannot be assigned to an irreducible 
representation of the symmetry group of the interaction.

The plan of the paper is as follows: in the next section, the 
``mixed density" prescription will be briefly discussed as a way to justify the introduction of the
projected density prescription to be discussed in Section 3. In Section 4 I will try to convey
what the open questions are regarding this issue as well as some thoughts about possible ways out. 
Conclusions will be given at the end of the paper. 

\section{Mixed density prescription}

In the evaluation of the matrix element of a density dependent ``Hamiltonian" between different 
states $|\varphi\rangle$ and $|\varphi '\rangle$ a prescription for the density dependent part 
is needed. A common choice, proposed in \cite{Bonche.90}  is the ``mixed density" prescription that
uses the overlap 
\begin{equation}
\rho_\textrm{Mixed} (\vec{r}) =
\langle \varphi | \hat{\rho} (\vec{r}) | \varphi ' \rangle /\langle \varphi | \varphi ' \rangle
\label{Eq:Mixed}
\end{equation}
as the density of the density dependent term of the ``Hamiltonian". In this
definition  $\hat{\rho} (\vec{r})=\sum_{i=1}^A \delta (\vec{r}-\vec{r}_i)$ is the density
operator. This prescription 
is based on the expressions obtained  with three body forces in spin saturated systems. The use of this
prescription makes the density dependent ``Hamiltonian" to break the symmetries broken by the 
density and also makes it a complex (non-hermitian) quantity.  
One might think, after considering these two
properties, that the mixed density prescription should be sentenced to oblivion. However, as 
was discussed in detail in Ref. \cite{RRRG.02}, the energy, which is the physical quantity to be 
computed out of the  above mentioned matrix elements, is a well defined quantity, i.e. it
is invariant under  the broken symmetries and it is a real quantity. To understand these facts
let us consider briefly the case of rotational invariance.  Let us now consider an intrinsic 
wave function $|\varphi\rangle$ that represents a deformed state and therefore is not 
characterized by the angular momentum quantum numbers. For standard (density independent)
Hamiltonians the energy is independent of the orientation
\begin{equation}
\langle \varphi | \hat{R}^+(\Omega) \hat{H} \hat{R} (\Omega) |\varphi\rangle=
\langle \varphi | \hat{H} |\varphi\rangle \label{Eq:Einv}
\end{equation}
as a consequence of the scalar nature of the Hamiltonian, 
namely $\hat{R}(\Omega) \hat{H} = \hat{H}\hat{R}(\Omega)$ for all Euler angles $\Omega$.
For density dependent ``Hamiltonians" $\hat{H}[\rho_{|\varphi\rangle}]$  it
was shown in Ref \cite{RRRG.02} what the scalar property for overlaps below holds
\begin{equation}
\langle\varphi|\hat{R}(\Omega) \hat{H}[\mbox{}_{R} \rho]|\varphi '\rangle = \langle\varphi|\hat{H}[\rho_R]\hat{R}(\Omega)|\varphi '\rangle
\label{Eq:EDDinv}
\end{equation}
where the shorthand notation 
$\rho_R=\langle\varphi|\hat{\rho}\hat{R}(\Omega)|\varphi'\rangle /\langle\varphi|\hat{R}(\Omega)|\varphi '\rangle$
and 
$\mbox{}_{R} \rho=\langle\varphi|\hat{R}(\Omega)\hat{\rho}|\varphi '\rangle /\langle\varphi|\hat{R}(\Omega)|\varphi'\rangle$ is used.
This property allows to extend Eq. (\ref{Eq:Einv}) to the case of density dependent 
``Hamiltonians" showing that the ensuing energy is orientation independent.

Concerning the reality of the energy, one has to realize that whenever a term like
$\langle\varphi| \hat{H}[\mbox{}_\varphi \rho_{\varphi'}]|\varphi '\rangle $
appears in the expression of the energy,  the symmetric term
$\langle\varphi'| \hat{H}[\mbox{}_{\varphi'} \rho_{\varphi}]|\varphi\rangle $
will also be present and weighted with the complex conjugate weight. 
It is not difficult to show \cite{RRRG.02} that
the last term is the complex conjugate of the first and therefore the contribution
of the two add up to a real quantity. This analysis is kind of a n\"aive one as the
density dependence in the Hamiltonian usually comes through a non integer power of the
(complex) density what forces the introduction of "branch cuts" in the complex plane
and to specify in which Riemann sheet the roots are defined. As a consequence
of this difficulty it was shown in Ref, \cite{Duguet.09} (see also
\cite{Dobaczewski.07} for an early consideration of this issue) that the energy was not
independent of the specific way it was computed. To be more specific, in the case of
PNP with density dependent ``Hamiltonians" consisting of 
a density dependent term with a density raised to a non integer power, the projected
energy was dependent upon the integration path chosen in the complex plane to perform
the integrations needed in the particle number procedure.  This is 
a real challenge to existing theoretical models as most of the Gogny, Skyrme or relativistic
models contain non-integer powers of the density in their definition. Recently,
an exception has shown up in the market in the form of the so called Barcelona-Catania-Paris
(BCP) functionals \cite{BCP,Robledo.08} which are only defined in terms of integer
powers of the density and are therefore promising candidates to incorporate the 
self-energy renormalization techniques of \cite{Lacroix.09} to produce well defined
quantities in calculations beyond the mean field.

A possibility that also comes up to solve the complex mixed density dilemma is to
use the so called ``projected density" prescriptions discussed next. 

\section{Projected density prescription}

The projected density prescription was postulated for Particle Number Projection (PNP) 
\cite{Valor.97,Valor.00} and Angular Momentum Projection (AMP) \cite{Schmid.04}
to solve, in one hand , the problems associated with the complex nature of the mixed density
prescription and, in the other, to give back to the density dependent part of the
``Hamiltonian" the nice property of preserving fundamental symmetries like rotational 
invariance. In this way, the density dependent part of the ``Hamiltonian" could be looked 
at as an effective
part of a real Hamiltonian and not as a device to generate a non trivial density dependence in
the energy functional \cite{Schmid.04}. 

In the case of PNP the particle number projected density 
\begin{equation}
\rho_{\textrm{N}}(\vec{r})=\frac{\langle\phi|\hat{P}_{N}^{+}\hat{\rho}(\vec{r})\hat{P}_{N}|\phi\rangle}
{\langle\phi|\hat{P}_{N}^{+}\hat{P}_{N}|\phi\rangle}
\end{equation}
was  considered in \cite{Valor.97,Valor.00} and it was even proved \cite{Valor.00} that this 
prescription as well as the mixed density prescription produced the same expressions for
the Lipkin Nogami parameter in the strong deformation limit. This has been the prescription
used in all particle number projected calculations with the Gogny force. In the rotational 
symmetry case the rotational invariant density corresponding to the $J=0$ projected state
$\rho_{\textrm{J=0}}(r)$ was postulated \cite{Schmid.04} to be used in the density 
dependent part of the Hamiltonian.  

The ``projected density" prescription looked  reasonable and the only argument found 
against it in the PNP case \cite{Robledo.07} was in the context of the GCM where, 
under some assumptions, it is possible to derive the Random Phase Approximation 
(RPA) as a small amplitude approximation of the GCM \cite{Jancovici.64}. 
This idea can be extended to the case
of density dependent interactions \cite{Robledo.07,Robledo.10} with the 
``mixed density" prescription to produce the RPA with the same matrix elements 
that appear in the standard derivation of the RPA for density dependent 
forces \cite{Blaizot.77}. However, this is not the case for the ``projected 
density" prescription
or other prescriptions postulated in the literature \cite{Duguet.03}.
It has also to be mentioned that the ``projected density" prescription produces in the 
context of the PNP and in the strong deformation limit an approximate 
expression for the projected energy (Kamlah expansion, see \cite{RS.80} for a detailed 
discussion) including a parameter, to be identified later with the Lagrange multiplier of
the particle number constraint,  that includes the rearrangement term 
characteristic of mean field theories with density dependent forces 
\cite{Valor.97,Valor.00,Robledo.07}. This property do not hold, however, for 
spatial symmetries like rotations as the leading term of the Kamlah expansion
for the projected density is not the intrinsic density as in the case of PNP but
some kind of symmetry preserving average (see below for the parity symmetry).
Therefore, in the general case the density dependent interaction of the underlying 
mean field is different from the density dependent interaction to be used in projection 
(i.e. both densities are ``decoupled" in a sense to make clear below) suggesting the
possibility to use the ``projected density" prescription 
also at the mean field level. That would make the Hartree-Fock-Bogoliubov (HFB) equations 
more involved due to the new rearrangement term but no fundamental problems were 
foreseen. As I will show below this possibility is also doomed to fail.

\begin{figure}
\begin{center}
\includegraphics[width=0.48\columnwidth]{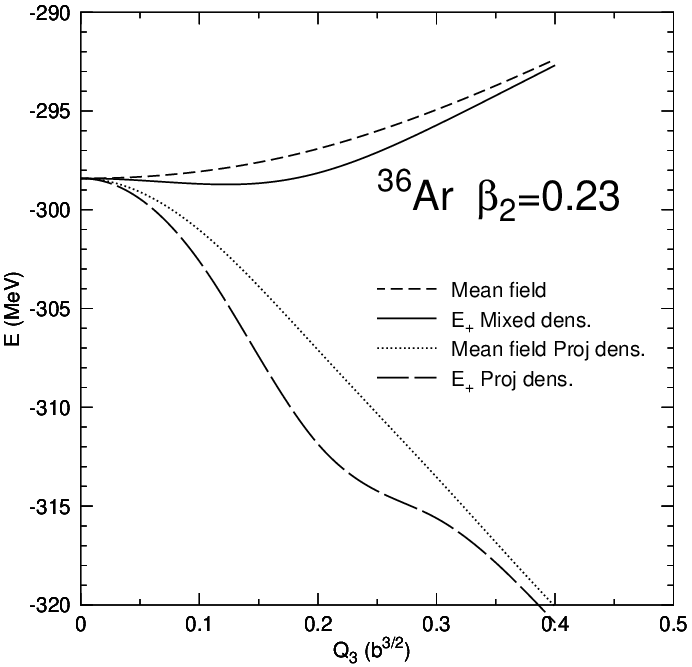}%
\includegraphics[width=0.48\columnwidth]{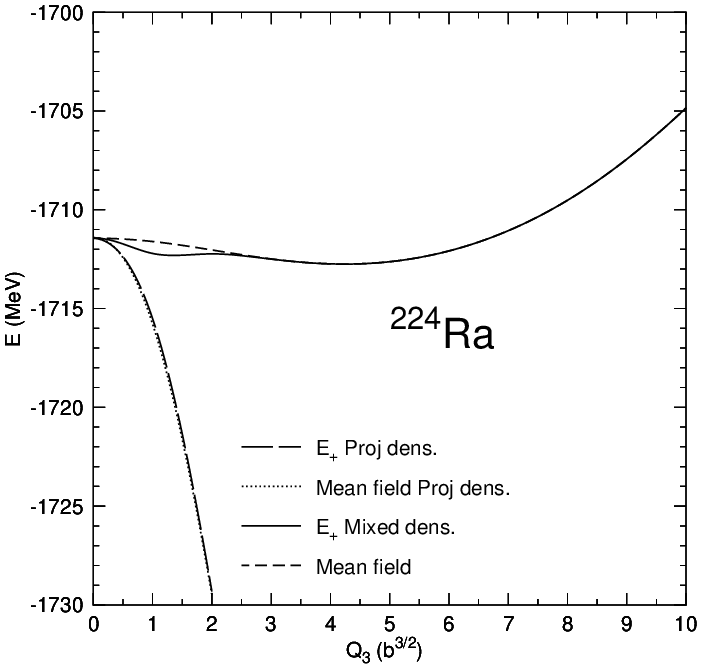}
\end{center}
\caption{(Color online) On the left hand side 
various energies for the nucleus $^{36}$Ar at a fixed value of $\beta_2$ of 
0.23 are plotted as a function of the octupole moment $Q_3$. The blue
dashed increasing curve corresponds to the intrinsic HFB energy. 
The red full line is for the projected energy to positive parity using
the ``mixed density" prescription. The dotted green down-sloping curve
is for a calculation of the HFB energy where the intrinsic density of the
density dependent interaction is replaced by the projected density
to positive parity. Finally, the purple full line down-sloping curve
is for the projected energy corresponding to positive parity and
with the parity projected density in the density dependent part of the 
interaction. The result was obtained with the Gogny D1S force.
On the right hand side, I plot the same quantities as before
but for the heavy actinide and octupole deformed nucleus $^{224}$Ra.
\label{Fig:Ener}}
\end{figure}

In order to show the catastrophic consequences of having in the density dependent
part of the ``Hamiltonian" a quantity that is decoupled from the intrinsic density
I will consider the case of parity projection of octupole deformed intrinsic
states. The Gogny \cite{Decharge.80} force D1S \cite{D1S} will be used for the 
calculations and I will proceed along
the same lines as our early study of parity projection with the Gogny force of
Refs. \cite{Egido.91,Garrote.98}. The parity projection operator to parity $\pi=\pm 1$ is
given by
\begin{equation}
\hat{P}_{\pi}=\frac{1}{\sqrt{2}}({I}+\pi\hat{\Pi})
\end{equation}
where $\hat{\Pi}$ is the parity operator. The projected energy is given by
\begin{equation}
E_\pi=\frac{\langle\phi|\hat{H} \hat{P}_\pi|\phi\rangle}{\langle\phi|\hat{P}_\pi|\phi\rangle}=\frac{\langle\phi| \hat{H} |\phi\rangle + 
                \pi\langle\phi| \hat{H}\hat{\Pi}|\phi\rangle}
                {\langle\phi|\phi\rangle+\pi\langle\phi|\hat{\Pi}|\phi\rangle
                }
\label{Eq:ProjE}
\end{equation}
The definite expression for the projected energy depends upon the prescription 
used for the density dependent part of the ``Hamiltonian".
For the ``mixed density" prescription the standard intrinsic density
$\rho (\vec{r})$ has to be used in the density dependent part of the Hamiltonian 
for  the evaluation of its mean value (first term in the numerator of Eq. (\ref{Eq:ProjE}) )
and the density $\theta(\vec{r})=\frac{\langle\phi|\hat{\rho}(\vec{r})
\hat{\Pi}|\phi\rangle}{\langle\phi|\hat{\Pi}|\phi\rangle}$ 
in the evaluation of 
$\langle\phi| \hat{H}\hat{\Pi}|\phi\rangle/\langle\phi|\hat{\Pi}|\phi\rangle$. 
In the case of the ``projected density" prescription the density 
\begin{equation}
\rho_{\textrm{PROJ}}^{\pi=+1}(\vec{r})=\frac{\langle\phi|\hat{P}_{\pi}^{+}\hat{\rho}
(\vec{r})\hat{P}_{\pi}|\phi\rangle}{\langle\phi|\hat{P}_{\pi}^{+}\hat{P}_{\pi}|\phi\rangle}
=\frac{\bar{\rho}(\vec{r})+\pi\langle\phi|\hat{\Pi}|\phi\rangle\bar{\theta}(\vec{r})}
{\left(1+\pi\langle\phi|\hat{\Pi}|\phi\rangle\right)}
\end{equation}
will be used in the density dependent part of the ``Hamiltonian". In this expression 
the averaged matrices $\bar{\rho}(\vec{r})=\frac{1}{2}\left(\rho(\vec{r})+\rho(-\vec{r})\right)$
and $\bar{\theta}(\vec{r})=\frac{1}{2}\left(\theta(\vec{r})+\theta(-\vec{r})\right)$ have been 
introduced.
In the strong deformation limit (corresponding to $\langle \phi |\hat{\Pi}|\phi\rangle\rightarrow 0$)
the projected density goes to the average density $\bar{\rho}(\vec{r})$ instead of the intrinsic one.
This is the fact that is at the heart of the problems encountered below for the projected energy. 

The positive parity projected energy $E_+$ has been computed using the above formulas and
prescriptions for given intrinsic configurations with different octupole moments and in
different nuclei. A light nucleus $^{36}$Ar has been chosen as an example of a system
with a density which is mainly surface and also as a system where the intrinsic ground state
is reflection symmetric (i.e. positive parity in this case). On the other hand, the heavy
nucleus $^{224}$Ra has been chosen as an example of an octupole deformed nucleus and with
a density which is mostly volume.  
The results obtained are shown in Fig. \ref{Fig:Ener} as a function of the octupole moment $Q_3$.
Several energies are computed: the mean field energy computed in the standard way $E$, the mean field
energy computed with the parity projected density in the density dependent ``Hamiltonian" 
$E_\textrm{PD}$ (the modified HFB framework discussed above), 
the positive parity projected energy computed with the ``mixed density" 
prescription $E_{+\,\textrm{MD}}$ and finally the positive parity projected energy 
computed with the ``projected density" prescription $E_{+\,\textrm{PD}}$. First, it is observed
the typical parabolic behavior of $E$ as a function of $Q_3$ for a non octupole deformed 
nucleus like $^{36}$Ar as well as the presence of a minimum for the case of the octupole 
deformed $^{224}$Ra case. The positive parity projected energy  $E_{+\,\textrm{MD}}$
obtained with the ``mixed density" prescription has a behavior close to the mean field energy
being lower than it in all the cases (see \cite{Egido.91}) and quickly converging to the mean field
energy for increasing values of the octupole moment. This is  a consequence of the behavior of the 
mean value of the parity that quickly decreases to zero as the octupole moment increases reaching
the strong deformation limit. 
On the other hand, the two corresponding energies
computed with the ``projected density" prescription show a very pronounced decrease of tens
of MeV that keep going as the octupole moment is increased. In these two cases, the 
configuration minimizing the corresponding energies would be an infinitely  deformed
octupole configuration with energies going to minus infinity. Obviously this is a catastrophic
situation that will lead the system to unphysical energies and intrinsic configurations.
Therefore the ``projected density" prescription, at least for spatial symmetries (see below), 
can be categorized as unphysical and its use has to be avoided.  Concerning the particle
number projected density prescription widely used with the Gogny force, the reasonable
results obtained so far clearly demonstrate that no catastrophe is taking place in that case.
However, and from a fundamental (as opposed to phenomenological) point of view,  the use of a 
prescription which yields to catastrophic results when used in other contexts should also be 
avoided. 

\begin{figure}
\begin{center}
\includegraphics[width=0.45\columnwidth]{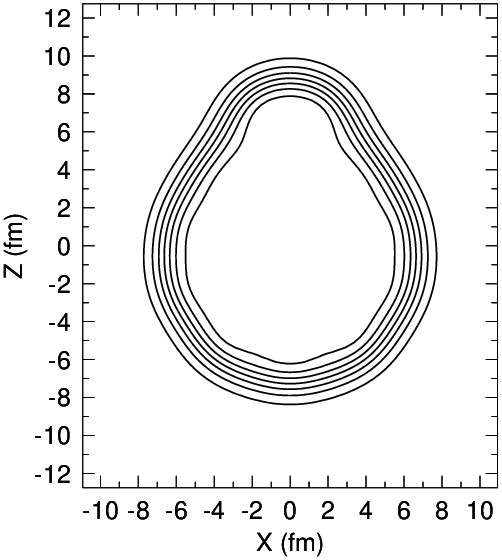}%
\includegraphics[width=0.45\columnwidth]{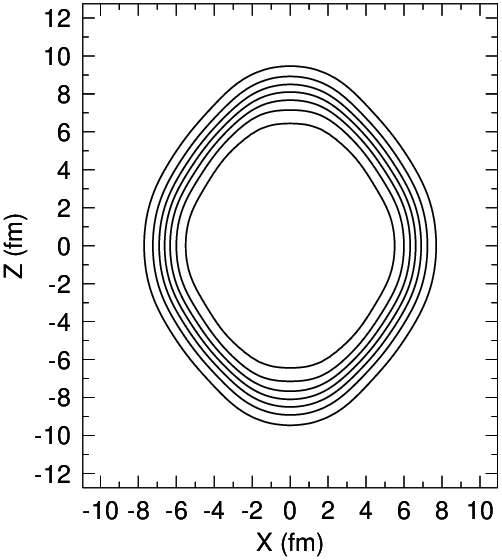}
\end{center}
\caption{On the left hand side, the contour plots of the intrinsic
density of the nucleus $^{224}$Ra with an octupole moment of
4 $b^ {3/2}$ (roughly corresponding to is ground state) are
depicted. The contour lines start at $\rho=0.02\textrm{fm}^{-3}$
for the outer contour and end at $\rho=0.14\textrm{fm}^{-3}$
for the inner one.  On the right hand side, contour plots
of the positive parity projected density corresponding to 
the intrinsic wave function of the left hand side plot. The 
contours are the same as in the left hand side plot. \label{Fig:Cont}}
\end{figure}

In order to understand the catastrophic results obtained it is convenient
to analyze the structure of the density dependent energy which is given
in the case of the Gogny force by the following integral
\begin{equation}
\langle \varphi | \hat{H}_{DD} | \varphi \rangle = \frac{t_3}{2} \sum_{\tau
,\tau '}(1+\frac{x_0}{2}-(x_0+\frac{1}{2})\delta_{\tau, \tau '}) \int 
d^ 3\vec{r} \rho_\tau \rho_{\tau '} \rho^\alpha
\end{equation}
where $\rho_\tau$ are the intrinsic densities with isospin $\tau$.
The replacement of the standard $\rho^\alpha(\vec{r})$ by 
$\rho_\textrm{Proj}^\alpha(\vec{r})$ produces a ``mismatch" in
the integrand because the new density $\rho_\textrm{Proj}$ does
not have the same shape as $\rho$ as can be seen in Fig. \ref{Fig:Cont} where
contour plots of the two densities are plotted for the nucleus $^ {224}$Ra at
an octupole moment $Q_3=4$b$^{3/2}$ (which corresponds to the ground state 
of this octupole deformed nucleus). As $\rho_\textrm{Proj}$
is reflection invariant it must have ``humps" at each side of the nucleus
whereas the intrinsic density only has a ``hump" at one of the sides (which
one depends on the sign of the octupole moment). The two ``humps" of the 
projected density will have half the density of the one of the intrinsic
density as the projected density is roughly given (strong deformation limit) by the average
density  $\bar{\rho}(\vec{r})=\frac{1}{2}\left(\rho(\vec{r})+\rho(-\vec{r})\right)$ 
(see Eq. (\ref{Eq:ProjE})). 
As a consequence of the mismatch of the different factors, the
integrand  corresponding to the ``projected density" prescription is slightly reduced 
as compared to the integrand with the intrinsic density in the density
dependent term. As a consequence, the value of the integral
is also reduced producing a reduction of the (repulsive) density
dependent energy that is magnified by the large values of the density dependent strength 
parameter $t_3$ ( 1562.22 MeV fm$^4$, 1609.46 MeV fm$^4$ and 1350.0 MeV fm$^4$ for 
Gogny D1M \cite{D1M}, D1N\cite {D1N} and D1S \cite{D1S}, respectively).

The mismatch between the projected density and the intrinsic density will also be 
very pronounced in the case of quadrupole deformation and associated to rotations
in space as well as in the case of spatial translations. It is therefore evident that
those two cases will also be plagued with the catastrophic decrease of the projected
energy for infinitely deformed configurations. Therefore in the case of angular momentum
projection and restoration of translational invariance the ``projected density" prescription
is doomed to badly fail.

\section{Open problems}
From the results presented here it is clear that, at least for spatial symmetries,
the only possibility to get rid of all the different problems encountered
in the extension of the density dependent forces paradigm to 
calculations going beyond mean field is to solely consider  forces/functionals depending
upon integer powers of the density. Apart from some scarce versions of Skyrme,
there is  a recently proposed functional with this property, the Barcelona-Catania-
Paris (BCP) functional \cite{BCP,Robledo.08} (see also an article in this issue by Baldo et
al concerning this functional \cite{Baldo.10}) which is based on a fit to a realistic 
nuclear matter equation of state followed by a Local Density Approximation map to produce
a functional for finite nuclei. Additional finite range surface terms, spin-orbit, Coulomb
and density dependent pairing are ingredients of this functional. Given the good performance
of BCP at the mean field level \cite{BCP,Robledo.08,Baldo.10}, this functional can become 
a serious contender for beyond mean field studies in the spirit of the Energy Density 
Functional (EDF).

Another alternative to get rid of the above problems would be to use 
approaches based on the Gaussian Overlap
Approximation (GOA) \cite{RS.80} or its more complete accurate and satisfactory version 
(the Topological GOA, see for instance \cite{Hagino.03} for a recent treatment) 
to perform the symmetry restorations and/or configuration mixing calculations (in the GCM 
spirit). This possibility has been explicitly suggested in \cite{Niksic.09} in the framework
of the Bohr Hamiltonian method but the philosophy behind was also implicit in early 
applications of the method \cite{Libert.99}. The reason why this proposal will cure the 
problems discussed before is because the residual interaction is defined as second derivatives
of the energy (where the density dependent term is well defined) and therefore are free from
the above problems of divergences and complex valued
quantities. Obviously, these ideas can be generalized to a framework where overlaps
are computed exactly but the residual interaction is defined as a second derivative 
in a way that closely resembles the RPA for density dependent 
forces/functionals. Work along this direction is in progress \cite{Robledo.10}.

\section*{Conclusions}
I have explicitly shown with an example using parity projection that the
``projected density" prescription, advocated by some authors to overcome
some difficulties encountered in the application of the ``mixed density" 
prescription in applications beyond the mean field, leads to catastrophic 
consequences for the projected
energy as this prescription favors a system where the reflection symmetry
violation is maximal and the projected energy in unbounded by below. This
consideration can also be extended to other symmetry restorations involving
spatial symmetry transformations like rotations and translations. 
The result is the consequence of the mismatch between the shape of the
intrinsic and the projected densities and the strongly repulsive character
of the density dependent term. 

\section*{Acknowledgments}
Work supported in part by the MICINN
(Spain) contract FPA2007-66069 and the Consolider-Ingenio 
2010 program CPAN (Spain) contract CSD2007-00042.

\end{document}